\documentclass{article}

\newcommand{\re}{\mathrm{e}}
\newcommand{\ri}{\mathrm{i}}
\newcommand{\rd}{\mathrm{d}}
\newcommand{\const}{\mathrm{const}}
\usepackage{latexsym}
\usepackage{graphics}
\usepackage{amsfonts,amssymb}

\def\openone{\leavevmode\hbox{\small1\kern-0.355em\normalsize1}}

\def\bbbc{{\mathbb C}}

\def\bbbr{{\mathbb R}}

\begin{document}
\noindent {\sf \Large \bf Superintegrability in the Manev Problem
and \\[5pt] its Real Form Dynamics}

\medskip
\bigskip
\noindent {\large \bf\sf V. S. Gerdjikov$^1$, A. Kyuldjiev$^1$, G.
Marmo$^2$, G. Vilasi$^3$} \medskip

\noindent {\sl $^1$Institute for Nuclear Research and Nuclear
Energy,\\72 Tzarigradsko chauss\'ee, 1784 Sofia, Bulgaria}
\medskip

\noindent {\sl $^2$Dipartimento di Scienze Fisiche, Universit\`a
Federico II di Napoli\\and INFN, Sezione di Napoli, Compl.\
Universitario di Monte Sant'Angelo,\\Via Cintia, 80126 Napoli,
Italy}
\medskip

\noindent {\sl $^3$Dipartimento di Fisica ``E.R. Caianiello",
Universita di Salerno,\\and INFN, Gruppo Collegato di Salerno,
Salerno, Italy}
\medskip

\noindent E-mails: gerjikov@inrne.bas.bg, kyuljiev@inrne.bas.bg,
marmo@na.infn.it, \\ \hspace*{1.3cm} vilasi@sa.infn.it


\begin{abstract}
We report here the existence of Ermanno-Bernoulli type invariants
for the Manev model dynamics which may be viewed upon as remnants
of Laplace-Runge-Lenz vector whose conservation is characteristic
of the Kepler model. If the orbits are bounded these invariants
exist only when a certain rationality condition is met and thus we
have superintegrability only on a subset of initial values. We
analyze real form dynamics of the Manev model and derive that it
is always superintegrable. We also discuss the symmetry algebras
of the Manev model and its real Hamiltonian form.
\end{abstract}

\section{Introduction}

Since time immemorial the circular motion was the archetype motion
of the heavenly bodies, and circle was assumed to be embodiment of
perfection---in the East and the West alike. No wonder that when
the observation data challenged this view a superposition of
several circular motions or circular motion with an off-centre Sun
were proposed in order to keep the circular paradigm intact.

Since Kepler and Newton elliptical trajectories replaced circular
ones as an archetype of the (bounded) planetary motion and the
circle is nowadays viewed upon\, rather as a degenerate ellipse
than as an ideal incarnated. The archetype of
elliptical motion is even exported to the atomic realm as we 
see depictions of\, `the atom' with ellipses representing
electrons' motion around the nuclei.

The advent of Einstein's theory did not produce a new archetype of
heavenly motions, apart from the exceptional case of a collapse
into the (still hypothetical) black holes. Nevertheless, among the
variety of relativistic effects the perihelion advance of inner
planets is definitely the best recognizable effect in the Solar
system. Maybe it is time to accept a new archetype:
{\em{precessing ellipse}}\/ (or more generally, precessing
conics). There are also classical arguments in its favour:
Kepler-type motion is generally not preserved by small
perturbations and generally any sort of `real world' interactions
like Solar pressure, drag etc would destroy `fixed ellipse' motion
\cite{Lacomba}. If precessing conics give us `the typical' motion
of planets it is tempting to ask which central force field
produces them. Surprisingly or not, the answer is: {\em{the Manev
model}} \cite{CBlaga}. Here we already have persistent KAM tori
and cylinders for a large class of even non Hamiltonian
perturbations \cite{Lacomba} and this is an additional argument in
favour of the Manev model.

Kepler problem is famous as one of archetypes of superintegrable
systems and probably, the first one where an unexpected
non-N\"other symmetry has been uncovered. It is intriguing to ask
whether Manev problem shares this property and here we report that
this is indeed the case, but not for all initial data. Artificial
models presenting such a behaviour are already known, e.g.\ sum of
squares of Hamiltonians of independent oscillators, but Manev
model is a `real world' example having this property. Let's remark
that for a generic central potential we could have disjoint set of
initial data corresponding to closed orbits but in our case
\emph{all} points on certain level sets of the angular momentum
lie on closed orbits which are intersections with the level sets
of the additional invariant.

Also we will show that real form dynamics of the Manev problem---a
closely connected dynamical model which we shall introduce
below---is superintegrable for \textsl{all}\/ initial data. Real
form dynamics of the Manev problem is interesting enough to
validate a separate analysis and we will describe it at the end of
the article.

\section{The Manev Problem Basics}

By Manev model \cite{Manev} we mean here the dynamics given by the
Hamiltonian:
\begin{equation}\label{H}
H=\frac{1}{2}(p_x^2 +p_y^2 +p_z^2) -\frac{A}{r} -\frac{B}{r^2}
\end{equation} where
$r=\sqrt{x^2+y^2+z^2}\,$; $A$ and $B$ are assumed to be arbitrary
real constants whose positive values correspond to attractive
forces. The genuine model proposed by George Manev was not
invented as an approximation of relativity theory but as a
consequence of Max Planck's (more general) action-reaction
principle and it derived a specific value for the constant
$B=\frac{3G}{2c^2}A$. Nevertheless, Manev model offers a
surprisingly good practical approximation to Einstein's
relativistic dynamics---at least at a solar system level---capable
to describe both the perihelion advance of the inner planets and
the Moon's perigee motion. In the last decade it had enjoyed an
increased interest either as a very suitable approximation from
astronomers' point of view or as a toy model for applying
different techniques of the modern mechanics (see e.g.\
\cite{MiocCR,Mioc,ManevFlow,Caballero,Diacu}).

Due to Hamiltonian's rotational invariance each component of the
angular momentum 
\begin{equation}\label{eq:L}
L_j = \varepsilon_{jkm}\,p_k x_m \qquad \{L_j\, ,
L_k\}=\varepsilon _{jkm}\,L_m \qquad \mbox{with  }\;
(x_1,x_2,x_3)=(x,y,z)
\end{equation}
is an obvious first integral: $\{H,\,L_j\}=0$ and so, like the
Kepler problem, the Manev model is integrable.
Components themselves are not in involution but span an $so(3)$
algebra with respect to the Poisson brackets.

The dynamics is confined on a plane which we assume to be $Oxy$
and is separable in radial coordinates. On the reduced phase space
(see e.g.\ \cite{GLMV} for the generalities of the reduction
procedure) obtained by fixing the angular momentum $L_z\equiv L$
to a certain value $\ell$ the motion is governed by:
\begin{equation}
H_{\rm eff}=\frac{1}{2}\left(p_r^2 +\frac{\ell^2-2B}{r^2}\right)
-\frac{A}{r}.
\end{equation}
The dynamics behave like radial motion of Kepler dynamics with
angular momentum squared $\ell^2-2B$; while the case $2B>\ell^2$
corresponds to overall centripetal effect. On the other hand, the
angular equation of motion $\dot{\theta}=L/r^2$ is still governed
by the `authentic' angular momentum $\ell$ (and $r$ is as just
described). Consequently, the remarkable properties of Kepler
dynamics that all negative energy orbits are closed and the
frequencies of radial and angular motions coincide (for any
initial conditions) are no more true. Thus we may have not only
purely classical perihelion shifts but also\, if $\,2B\ge\ell^2\ne
0\,$ we may have collapsing trajectories which are spirals, even
though in phase space they are symplectic transformations; while
in the Kepler dynamics the only allowed fall down is along
straight lines. For this reason the set of initial data leading to
collision has a positive measure and this may offer an explanation
why collisions in the solar system are estimated to happen more
often than Newton theory predicts \cite{Spiral}.

The dynamics of Manev model has already been thoroughly analyzed
(see e.g. \cite{MiocCR,Mioc,ManevFlow,Caballero,Diacu}) and we
shall concentrate here on some of its invariants and resulting
symmetry algebra.

\section{The Kepler Problem Invariants}\label{ssec:2.1}

In the case of Kepler problem, corresponding to $B=0$, we have
more first integrals (for details and historical notes see e.g.
\cite{Nucci,Leach,Vilasi,Antonella}):
\begin{equation}\label{eq:LaRuLe}
J_x=p_yL + \frac{A}{r}x\,, \qquad J_y=-p_xL +
\frac{A}{r}y\,,\qquad \{H_K,\vec{J}\}=0.
\end{equation}
where 
$H_K$ is the Kepler Hamiltonian and $J_x $ and $J_y $ are the
components of the Laplace-Runge-Lenz vector. They are not
independent since
\begin{equation}\label{J^2}
J^2=2H_K L^2+A^2.
\end{equation}
Together with the Hamiltonian and angular momentum they close on
an algebra with respect to the Poisson brackets:
\begin{equation}\label{eq:PB-alg1}
\{H_K,L\}=0\,,\;\;\{L,J_x\}=J_y\,,\;\;\{L,J_y\}=-J_x\,,
\;\;\{J_x,J_y\}=-2H_K L.
\end{equation}
After redefining $\vec{E}={\vec{J}}/{\sqrt{|-2H_K|}}$ we get:
\begin{equation}\label{eq:PB-alg2}
\{L,E_x\}=E_y\,,\qquad\{L,E_y\}=-E_x\,,\qquad\{E_x,E_y\}=-{\mathrm
{sign}}(H_K)\,L
\end{equation}
which makes obvious the fact that we have an $so(3)$ algebra for
negative energies and $so(2,1)$ for positive ones. In the case of
the 3-dimensional Kepler problem the components of the angular
momentum give us another copy of $so(3)$, see eq. (\ref{eq:L}), so
the full symmetry algebra is $so(4)$ or $so(3,1)$ depending on the
sign of $H_K$.


According to \cite{Leach}, the first use of these first integrals
was made by J.\,Hermann (= J.\,Ermanno) in 1710 (in order to find
\emph{all} possible orbits under an inverse square law force) in
the disguise of `Ermanno-Bernoulli' constants:
\begin{equation}\label{eq:J_pm}
J_\pm = J_x \mp {\mathrm i} J_y=\left(\frac{L^2}{r}-A\,
 \mp\, {\rm i}\, Lp_r\, \right){\rm e}^{\pm \rm i\theta}
\end{equation}
which satisfy:
\begin{equation}\label{eq:PB-alg5}
\{H_K,J_\pm\}=0, \qquad\{L,J_\pm\}=\pm{\rm i} J_\pm,
\qquad\{J_+,J_-\}=-4{\rm i}H_K L.
\end{equation}

Curiously enough, the initial Kepler idea for a circular orbits
with an off-centre Sun happens to be correct in a different
context. As discovered by Hamilton \cite{Milnor}, the
\emph{velocity} vector in the Kepler dynamics moves along a circle
laying in a plane containing the origin but, in general, not
centered, at the origin. If we choose the $x$-axis pointing to the
point of closest approach, the centre of this circle is located at
$(0,\epsilon A /\ell,0)$ and its radius is $A/\ell$ where
$\epsilon=\sqrt{1+{2E\ell^2}/{A^2}}$ is the eccentricity of the
orbit. If the spatial orbit is hyperbola ($\epsilon >1$) the
velocity space orbit is only a section of the circle, otherwise
the full circle is traversed.

\section{The Manev Problem Invariants}\label{ssec:2.2}

Assuming $0\ne\ell^2>2B$ and denoting \begin{equation}
\nu^2=\frac{\ell^2-2B}{\ell^2}\,,\qquad w=\frac{L^2}{r}-
\frac{\ell^2}{\ell^2-2B}A
\end{equation}
one easily verifies that
\begin{equation}\label{Jdot=0}
\frac{\rd}{\rd \theta}\left[\left(\nu w \pm \ri
\frac{\rd}{\rd\theta}w\right)\re^{\pm\ri \nu \theta}\right]= 0
\end{equation}
and since $\frac{\rd}{\rd \theta}=\frac{r^2}{L}\frac{\rd}{\rd t}$
we obtain:
\begin{equation}\label{J1dot=0}
\frac{\rd}{\rd t}\left[\left(\nu \frac{L^2}{r} -\frac{A}{\nu} \mp
\ri Lp_r\right)\re^{\pm\ri \nu \theta}\right]= 0.
\end{equation}

In the case when $\ell\ne 0,\; \ell^2>2B\,,\;H<0$ and $A>0$ the
motion is on a 2-dimensional torus. In order to have globally
defined constants of motion in this case we have to require that
the $\nu$'s be rational i.e.\
\begin{equation}\label{ration}
\nu=\sqrt{\ell^2-2B}:\ell=m:k
\end{equation}
with $m$ and $k$ integers. Then due to eq. (\ref{Jdot=0})
\begin{equation}
{\mathcal J}_\pm= {\mathcal J}_\mp^*=\left[ {m\over k}{L^2\over
r}- {k\over m}A\, \mp\, {\rm i}\, Lp_r\, \right]{\rm e}^{\pm \rm i
m\theta /k}
\end{equation}
are conserved by the flow of eq.\,(\ref{H}) on a surface $L=\ell$
satisfying the rationality condition (\ref{ration}). Thus we have
conditional constants of motion which exist only for disjoint but
infinite set of values $\ell$ (c.f.\ invariant relations of
\cite{Levi-Civita}).

The trajectory in the configuration space is a `rosette' with $m$
petals and this is connected to the fact that ${\mathcal J}_\pm$
are invariant under the action of the cyclic group generated by
rotations by angle $\frac{2\pi k}{m}$:
\begin{equation}\label{theta+}
\theta\rightarrow\theta +2\pi\frac{k}{m}n \qquad n=0,1,\ldots ,
m-1\,.
\end{equation}
While in the Kepler case we could unambiguously attach the
Laplace-Runge-Lenz vector to Ermanno-Bernoulli invariants this is
not possible now due to this finite symmetry. (It is intuitively
clear that if the Laplace-Runge-Lenz vector points to the
perihelion of the Kepler ellipse, 
now we have $m$ petals to choose between.) Anyway, up to this
unambiguity, \emph{or} restricting ourselves to one of the $m$
sectors\, we may note that while the radial/angular components of
the Laplace-Runge-Lenz vector take the form:
\begin{equation}
J_r=\frac{L^2}{r}-A \qquad J_\theta =-Lp_r
\end{equation}
in our case $\;{\mathcal J}_r +{\rm i}{\mathcal
J}_\theta=\left(\nu\frac{\ell^2}{r}-\frac{A}{\nu}-{\rm i}\ell
p_r\right){\rm e}^{{\rm i}(\nu -1)\theta}$ and hence:
\begin{eqnarray}
{\mathcal
J}_r&=&\left(\nu\frac{\ell^2}{r}-\frac{A}{\nu}\right)\cos (\nu
-1)\theta + \ell p_r \sin (\nu -1)\theta \\{\mathcal J}_\theta
&=&-\ell p_r \cos (\nu -1)\theta
+\left(\nu\frac{\ell^2}{r}-\frac{A}{\nu}\right)\sin (\nu -1)\theta
\,.\nonumber
\end{eqnarray}
\vspace*{.35em}

Turning to the algebraic properties of the new invariants one
finds that the Poisson brackets between the real and imaginary
parts of ${\mathcal J}_\pm =\mathcal{J}_0\mp \ri
\mathcal{J}_1=\mathcal{J}_x\mp \ri \mathcal{J}_y$ are:
\begin{eqnarray}\label{eq:PB-alg26}
\{H,{\mathcal  J}_{0,1}\}&=&0, \qquad \{L,{\mathcal J}_0\}={m\over
k}{\mathcal J}_1, \qquad \{L,{\mathcal J}_1\}=-{m\over
k}{\mathcal J}_0\nonumber\\
\label{eq:PB-alg27} \{{\mathcal J}_0 ,{\mathcal  J}_1\}&=&-
{m\over k}L \left[2H
-\frac{2B}{r^2}\left(\frac{2L^2}{\ell^2}-1\right)\right].
\end{eqnarray}

Here we have a $1/r^2$ term which seems to obstruct the Poisson
brackets to form a closed algebra. Fortunately, redefining
$\mathcal{E}_{0,1}={\mathcal  J}_{0,1}/L{\sqrt{L^2-\ell^2}}$ we
obtain the closed algebra $\mathfrak{g}_{H,L} $:
\begin{eqnarray}\label{eq:PB-alg16}
\{H,{\mathcal E}_{0,1}\}&=&0,\qquad \{L,{\mathcal E}_0\}={m\over
k} {\mathcal E}_1, \qquad \{L,{\mathcal E}_1\}=-{m\over
k}{\mathcal E}_0
\nonumber\\
\label{eq:PB-alg17} \{{\mathcal E}_{0} ,{\mathcal E}_{1}\}&=&
{m\over k}\frac{1}{(L^2-\ell^2)^2} \left[2HL+
\frac{k^2}{m^2}\frac{A^2}{L^3} \frac{2L^2-\ell^2}{L^2-\ell^2}
\right]
\end{eqnarray}
in which $H $ is a central element and $L $, $\mathcal{E}_0 $ and
$\mathcal{E}_1 $ can be viewed as Cartan and root-vector
generators. Due to (\ref{eq:PB-alg16}) $\mathfrak{g}_{H,L} $ is a
deformation of $gl(2) $. Of course, we have in addition the
$so(3)$ algebra (\ref{eq:L}).

Similarly, in the case when $0\ne\ell^2<2B$ we may denote
$\frac{2B-\ell^2}{\ell^2}=\upsilon^2$ with $\upsilon$ real and
\begin{equation} {\mathcal E}_\pm=
\left[ \upsilon{L^2\over r}+ \frac{A}{\upsilon}\, \mp\, Lp_r\,
\right] { {\rm e}^{\mp \upsilon\theta} \over L \sqrt{L^2-\ell^2}}
\end{equation}
are first integrals for any $\ell$ and they satisfy:
\begin{eqnarray}\label{eq:PB-alg8}
\{H,{\mathcal E}_\pm\}&=&0\,,\qquad \{L,{\mathcal E}_\pm\}=\mp
\upsilon{\mathcal E}_\pm  \nonumber\\
\label{eq:PB-alg9} \{{\mathcal E}_+ ,{\mathcal E}_-\}&=&
\frac{2\upsilon}{(L^2-\ell^2)^2}\left[2HL -\frac{A^2}{\upsilon^2
L^3} \frac{2L^2-\ell^2}{L^2-\ell^2} \right].
\end{eqnarray}
The algebra $\mathfrak{g}_{H,L}'$ satisfied by $H $, $L $ and
$\mathcal{E}_{\pm} $ is quite analogous to $\mathfrak{g}_{H,L} $
but with a different function at the right hand side of the
bracket $\{\mathcal{E}_+,\mathcal{E}_-\} $.

Finally, when $\ell^2=2B$ we have the first integral:
\begin{equation}
j= Lp_r+A\theta
\end{equation}
satisfying $\{H,j\}=0$, $ \{L,j\}=A$. \vspace*{.5em}

\section{Real Form Dynamics}

Here we briefly recall the notion of real form (RF) dynamics
referring the reader to \cite{We} for more details and a list of
examples.

We start from a standard (real) Hamiltonian system ${\mathcal
H}\equiv \{{\mathcal M},\omega ,H \}$ with $n$ degrees of freedom
and at the present stage we assume that our phase space is just a
vector space ${\mathcal M}=\bbbr^{2n}$.

Let's consider its complexification: $ \mathcal{H}^\bbbc \equiv
\left\{ \mathcal{M}^\bbbc , H ^\bbbc , \omega^\bbbc \right\}$
where $\mathcal{M}^\bbbc $ can be viewed as a linear space over
the field of complex numbers:
\[ \mathcal{M}^\bbbc = \mathcal{M}\oplus \rm i \mathcal{M}.
\]
In other words the dynamical variables in $\mathcal{M}^\bbbc $ now
take complex values. We assume that the Hamiltonian $H$ (as well
as all other possible first integrals in involution $I_k$) are
{\em{real analytic functions}} on $\mathcal{M} $ which can
naturally be extended to $\mathcal{M}^\bbbc$. We introduce on the
phase space ${\mathcal M}$ an involutive, symplectic automorphism
${\mathcal  C} : {\mathcal M}\rightarrow {\mathcal M }$:
\begin{equation}\label{eq:C-inv1}
{\mathcal  C}^2=\openone \,, \qquad 
\mathcal{C} (\{ F,G\}) = \{ \mathcal{C}(F), \mathcal{C}(G)\}
\end{equation}
where with some abuse of terminology we use the same notation for
the action of ${\mathcal C}$ on the dual of the phase space.

Since ${\mathcal C}$ has eigenvalues $1$ and $-1$, it naturally
splits ${\mathcal M}$ into two eigenspaces:
\begin{equation}\label{eq:C-inv3}
{\mathcal M}={\mathcal M}_+\oplus {\mathcal M}_-
\end{equation}
whose dimensions need not be equal. Due to the fact that
${\mathcal C}$ is symplectic ${\mathcal M}_-$ and ${\mathcal M}_+$
are symplectic subspaces of ${\mathcal M}$ and we will write
$\omega = {\omega}_+ \oplus {\omega}_-.$

Assuming a symplectic frame adapted to ${\mathcal C}$ we have:
\[
\omega = \sum_{k=1}^{n_+} \rd p_{k+} \wedge \rd q_{k+} +
\sum_{k=1}^{n_-} \rd p_{k-} \wedge \rd q_{k-}.\] The automorphism
${\mathcal C}$ can naturally be extended to ${\mathcal M}^\bbbc$
and it splits it again into a direct sum of two eigenspaces:
$$
{\mathcal M }^{\bbbc} ={\mathcal M }^{\bbbc}_- \oplus {\mathcal M
}^{\bbbc}_+.
$$
Similarly, the action of the complex conjugation $^*$ produces
splitting into real and imaginary parts of the corresponding
spaces. By construction  ${\mathcal C}$ commutes with $*$ and
their composition $\,\widetilde{\mathcal C} \equiv {\mathcal
C}\,\circ\, ^* =\, ^*\circ\, {\mathcal C}\,$ is also an involutive
symplectic automorphism on ${\mathcal M}^\bbbc$; then we define
${\mathcal M}_{\bbbr}$ to be the fixed point set of
$\widetilde{\mathcal C}$ i.e.
$${\mathcal M}_{\bbbr}={\sf Re}{\mathcal M }^{\bbbc}_+\oplus
\rm i\,{\sf Im}{\mathcal M }^{\bbbc}_-$$ and it is again a
symplectic subspace. From now on we will be interested in dynamics
on ${\mathcal M}_{\bbbr}$ and its connection to the initial real
dynamical system.

In order to construct `real form dynamics' we shall assume that
the Hamiltonian is $\mathcal{C}$-invariant, i.e.:
\begin{equation}\label{eq:H-inv}
{\mathcal C}(H) = H.
\end{equation}
Then the Hamiltonian on the complexified phase space $H^{\bbbc}$
(being the same analytical function of the complexified variables)
will share this property.

The {\em `real form dynamics'} may be defined either as:
\begin{description}
\item i) complexified Hamilton equations on ${\mathcal M}^\bbbc$
being consistently restricted to ${\mathcal M}_{\bbbr}$. This
gives a real vector field tangent to ${\mathcal M}_{\bbbr}$ and
satisfying the equations of motion given by the real part of
$H^{\bbbc}$ {\emph {or}}
\item ii) dynamics on ${\mathcal M}_{\bbbr}$ defined by the
restricted $H^{\bbbc}$ and $\omega^\bbbc$ (whose restrictions are
real on ${\mathcal M}_{\bbbr}$):
\begin{eqnarray}\left.H\right| _{{\mathcal M }_{\bbbr}}\! &=&
{H+\widetilde{\mathcal C}(H)\over 2} = {H+{\mathcal C}(H)^*\over
2} = {\sf
Re} H^{\bbbc}\nonumber\\
\left.\omega^\bbbc\right| _{{\mathcal M }_{\bbbr}} &=& \rd\, {\sf
Re} p^\bbbc_+ \wedge \rd\, {\sf Re} q^\bbbc_+ - \rd\, {\sf Im}
p^\bbbc_- \wedge \rd\, {\sf Im} q^\bbbc_-.
\end{eqnarray}
\end{description}

Now we have a well defined dynamical system ${\mathcal H}_{\bbbr}
= \{{\mathcal M }_{\bbbr}$, $\left.\omega\right| _{{\!\mathcal M
}_{\bbbr}}$, $\left.H\right| _{{\mathcal M }_{\bbbr}}\}$ with {\em
real Hamiltonian} and {\em real symplectic form} on a subspace of
the complexified phase space.

It is noteworthy that the `real form dynamics' corresponding to a
Liouville integrable Hamiltonian system is Liouville integrable
again \cite{We}. Similarly, the `real form dynamics' corresponding
to a superintegrable Hamiltonian system is superintegrable again.
In such a case we have $\kappa \in [n+1,\, 2n-1]$ independent
constants of motion which are no more in involution. It could
easily be checked that they will again produce $\kappa$
independent constants of motion of the RF dynamics.

\section{Real Form Dynamics of the Manev Problem}

The Manev Hamiltonian (and the canonical symplectic form as well)
is invariant under the involution $\mathcal C$ reflecting the
$y$-degree of freedom:
\begin{eqnarray}\label{eq:C}
\mathcal{C} (x)=x, \qquad&\mathcal{C} (y)=-y,& \qquad \mathcal{C}
(z)=z \nonumber\\
\mathcal{C}(p_x)=p_x,\qquad&\mathcal{C} (p_y)=-p_y,&\qquad
\mathcal{C} (p_z)=p_z .
\end{eqnarray}
Consequently, the `real form dynamics' of Manev model for this
choice of involution will be given by:
\begin{eqnarray}\label{HRF}
H_{\bbbr}&=&{1\over 2}(p_x^2 -p_y^2 +p_z^2) -{A\over \rho}-{B\over
\rho^2}
\\\omega_{\bbbr}&=&\rd p_x\wedge \rd x - \rd p_y\wedge \rd
y + \rd p_z\wedge \rd z \nonumber
\end{eqnarray}
where $\rho=\sqrt{x^2-y^2 +z^2}$ is the `radius' of the
pseudo-sphere . This is not an ordinary central field dynamics but
rather an `indefinite metric central field' as $H_{\bbbr}$ depends
on indefinite metric distance $\rho$. The real form Hamiltonian
$H_{\bbbr}$ and the appropriate `angular momentum' $\tilde{L}_j$
are still commuting first integrals and the model is integrable.
The involution acts on $\tilde{L}_j$ according to:
$\,\mathcal{C}(\tilde{L}_j) = (-1)^{j} \tilde{L}_j \,$ and
\begin{equation}\label{eq:L-rf}
\{\tilde{L}_j\,, \tilde{L}_k\} =\varepsilon_{jki}(-1)^{j+k+1}
\tilde{L}_i
\end{equation}
instead of eq.\ (\ref{eq:L}); the corresponding algebra is
$so(2,1) $ which is the real form of $so(3)$ obtained with a
$\mathcal{C}$-induced Cartan involution.

We shall assume again that the motion is on the $Oxy$-plane and in
order to avoid the question of the behavior of trajectories on the
singularities we restrict our attention on the
$\mathcal{C}$-invariant configuration space:
$$\{(x,y,z)\in\mathbb{R}^2 \mid z=0, x^2>y^2, x>0\}.$$
Then the dynamics is separable in pseudo-radial coordinates
$\,\vartheta=\mathrm{artanh}(y/x) \in (-\infty, \infty)$ and $\rho
\in (0,\infty)$:
\begin{eqnarray}\label{HeffRF}
H&=&{1\over 2}\left(p_{\rho}^2 - {\pi_{\vartheta}^2 \over
\rho^2}\right) - {A\over \rho}-{B\over \rho^2}\\\omega &=&\rd
p_\rho\wedge \rd \rho + \rd \pi_\vartheta\wedge \rd
\vartheta\nonumber
\end{eqnarray} with
$\tilde{L} \equiv \tilde{L}_z=\pi_{\vartheta}$, hence
$\dot{\pi_{\vartheta}}=0$ and
$\dot{\vartheta}={-\tilde{L}/\rho^2}$. Due to the different
symplectic form $\tilde{L}$ generates now transformations which
preserve $\rho$.

The type of the $\rho$-trajectories could be easily read off after
the observation that the value of the real form Hamiltonian will
be:
$$h={1\over 2}\left(\dot{\rho}^2 - {\ell^2 \over \rho^2}\right) -
{A\over \rho}-{B\over \rho^2}$$ due to $\dot{x}^2-\dot{y}^2 =
\dot{\rho}^2 - {\ell^2 /\rho^2}$ and denoting the value of
$\tilde{L}$ by $\ell$. Introducing $v=\dot{\rho}$ and $u={1/\rho}$
we obtain an equation describing conics in the
$(u,v)$-space:$$u^2(\ell^2 + 2B) +2Au +(2h-v^2)=0.$$ Performing
the same type of analysis as in \cite{Mioc} we may conclude that
we may have three types of qualitatively different dynamical
regimes:

\begin{description}

\item for $\ell^2+2B>0$ we have a family of hyperbolas.

\item for $\ell^2+2B=0$ we have a family of parabolas for $A\ne 0$
which degenerate at $A=0$ into pair of lines parallel to the $1/
\rho$-axis.

\item for $\ell^2+2B<0$ (only possible for repulsive Manev term)
we have a family of ellipses. \vspace{-.4em}

\end{description}
Of course, in all these cases we have to exclude the region
$u<0$.\vspace{.35em}

In order to obtain more specific information about the motion we
will need an equation for the trajectories. Let's note again that
in the case of non-vanishing angular momentum we have
$dt=-\frac{\rho^2}{\tilde{L}}d\vartheta$. As a result the equation
for the $\rho$-motion takes the form:
\begin{equation}\label{RF_eq}
\frac{\rd^2}{\rd\vartheta^2}\frac{\tilde{L}^2}{\rho}-\frac{\ell^2+2B}{\ell^2}
\frac{\tilde{L}^2}{\rho}-A=0.
\end{equation}
Assuming $\ell^2+2B \ne 0$ we introduce \begin{equation}
\upsilon^2=\frac{\ell^2+2B}{\ell^2}\,,\qquad
w=\frac{\tilde{L}^2}{\rho}+ \frac{\ell^2}{\ell^2+2B}A
\end{equation}
and obtain an inverted oscillator--type equation:
\begin{equation}
\frac{\rd^2}{\rd\vartheta^2}w-\upsilon^2 w=0.
\end{equation}
Denoting by $c_j$ the integration constants below we conclude
that:
\begin{description}
\item If $\ell^2+2B>0$ the solution $\rho^{-1}(\vartheta)$ will be:
\begin{equation}
\frac{\tilde{L}^2}{\rho}=c_1 \cosh(\upsilon\vartheta) +c_2
\sinh(\upsilon\vartheta)-\frac{A}{\upsilon^2}.
\end{equation}
Trajectories may collapse ($\rho\rightarrow 0$) for
$\vartheta\rightarrow \pm \infty $ and $c_1>c_2>0$, {\em or}
$\rho$ may tend to $ \infty$ as $\vartheta$ tends to certain
values $\vartheta_{min}$ and $\vartheta_{max}$.

\item If $\ell^2+2B=0$ the solution of eq.\ (\ref{RF_eq}) will be:
\begin{equation}
\rho^{-1}=\frac{A}{2\tilde{L}^2}\vartheta^2 + c_3 \vartheta
+c_4\,.
\end{equation}
If $A>0$  trajectories collapse for $\vartheta\rightarrow \pm
\infty$ and if $A<0$ then $\rho\rightarrow \infty$ as $\vartheta$
tends to some $\vartheta_{min}$ and $\vartheta_{max}$. The case
$A=0$ leads to linear solution $\rho^{-1}(\vartheta)$ and
corresponds either to the motion along fixed $\rho$ or to
trajectory starting at $\rho=\infty$ and some value of $\vartheta$
and collapsing for $\vartheta\rightarrow \infty$ (or its reverse).

\item If $\ell^2+2B<0$ we will have a solution which oscillates
harmonically between some values $\rho_{min}$ and
$\rho_{max}$:\begin{equation} \frac{\tilde{L}^2}{\rho}=c_5
\cos(\upsilon\vartheta) +
c_6\sin(\upsilon\vartheta)-\frac{A}{\upsilon^2}.
\end{equation}
The case when $\rho_{min}<0$ means that `acceptable' motions will
be trajectories coming from $\rho=\infty$ at
$\vartheta=\vartheta_{min}$ and going to $\rho=\infty$ for some
$\vartheta=\vartheta_{max}$.\vspace{-.25em}

\end{description}

In the special case of vanishing angular momentum we have
$1$-dimensional motion along the ray $\vartheta=\const$. It may be
oscillating between some $\rho_{min}$ and $\rho_{max}$ or heading
to collapse, or to infinity.

It is worth noting that the only trajectories which are compact in
the  $(x,y)$-space are those collapsing at both their ends in the
origin tangentially to the boundaries $x=\pm\, y$ and those
oscillating on a line interval of $\vartheta=\const$. This is to
be contrasted to the standard Manev or Kepler problems.

Obviously, when $B=0\,$ we will obtain a real form dynamics of the
Kepler model. In this case we have even fewer possibilities for
compact trajectories as we could not have oscillations along the
line of $\vartheta=\const$.

Since the motion is never on a 2-torus the additional first
integrals are always globally defined for all initial data. When
$0\ne\ell^2>-2B\,$ they take the form:
\begin{equation}
 {\mathcal J}_\pm = \left[\upsilon
 {\tilde{L}^2\over \rho}+\frac{A}{\upsilon}\,\pm\, \tilde{L}p_{\rho}\,
\right] {\rm e}^{\mp \upsilon\vartheta}.
\end{equation}
As we did earlier we can introduce the renormalized
$\mathcal{E}_\pm =
\mathcal{J}_\pm/\tilde{L}\sqrt{\tilde{L}^2-\ell^2} $ and derive
for them the following symmetry algebra
$\mathfrak{g}_{H,\tilde{L}}' $
\begin{eqnarray}\label{eq:PB-alg29}
\{H_{\bbbr},{\mathcal E}_\pm\}&=&0, \qquad \{\tilde{L},{\mathcal
E}_\pm\}=\mp\upsilon{\mathcal E}_\pm \nonumber\\
 \{{\mathcal E}_{+} ,{\mathcal E}_{-}\}&=&
\frac{2\upsilon}{(\tilde{L}^2-\ell^2)^2} \left[2H_\bbbr \tilde{L}
-\frac{A^2}{\upsilon^2 \tilde{L}^3}
\frac{2\tilde{L}^2-\ell^2}{\tilde{L}^2-\ell^2}\right].
\end{eqnarray}
Like in (\ref{eq:PB-alg16}) above $\mathfrak{g}_{H,\tilde{L}}' $
is a deformation of $gl(2) $ having the same $H $, $\tilde{L} $
dependence in the right hand side of (\ref{eq:PB-alg29}), though
$L$ and $\tilde{L}$ have different properties.

Note that the algebras $\mathfrak{g}_{H,L} $ and
$\mathfrak{g}_{H,\tilde{L}}' $ seem very close, i.e.\ they do not
change effectively when passing from one real Hamiltonian form to
the other. The reason for this is the fact, that all its
generators are invariant with respect to the involution
$\mathcal{C} $. The situation changes when we consider the algebra
satisfied by $\tilde{L}_j \,$, see eq.~(\ref{eq:L-rf}).

In the case when $0\ne\ell^2<-2B\;$ let
$\,\nu^2=\frac{-(\ell^2+2B)}{\ell^2}\;$ and then invariants are:
\begin{equation}
{\mathcal J}_\pm= {\mathcal J}_\mp^*={\mathcal J}_0 \mp\ri
{\mathcal J}_1=\left[ \nu{\tilde{L}^2\over \rho}- \frac{A}{\nu}\,
\mp\, {\rm i}\, \tilde{L}p_\rho\, \right]{\rm e}^{\mp \rm i
\nu\vartheta}
\end{equation}
Redefining again: ${\mathcal E}_{0,1}={\mathcal
J}_{0,1}/\tilde{L}\sqrt{\tilde{L}^2-\ell^2}$ we obtain the
brackets:
\begin{eqnarray}\label{eq:PB-alg19}
\{H_{\bbbr},{\mathcal E}_{0,1}\}&=&0, \qquad \{\tilde{L},{\mathcal
E}_0\}=-\nu{\mathcal E}_1, \qquad \{\tilde{L},{\mathcal
E}_1\}=\nu{\mathcal E}_0\\
\{{\mathcal E}_{0} ,{\mathcal E}_{1}\}&=&
\frac{-\nu}{(\tilde{L}^2-\ell^2)^2} \left[2H_\bbbr \tilde{L}
+\frac{A^2}{\nu^2
\tilde{L}^3}\frac{2\tilde{L}^2-\ell^2}{\tilde{L}^2-\ell^2}
\right].\nonumber
\end{eqnarray}

Finally, when $\ell^2=2B$ we have the first integral:
\begin{equation}
j= \tilde{L}p_\rho-A\vartheta
\end{equation}
satisfying $\{H,j\}=0$, $ \{\tilde{L},j\}=-A$.

\section{Conclusions}\label{sec:C}

The discovered existence of Ermanno-Bernoulli type invariants
strengthens our belief that Manev model has an exceptional
position among the central field theories. Not only it provides a
better description of the real motion of the heavenly bodies than
Kepler model but to a large extent it shares its
superintegrability---probably its most celebrated mathematical
feature. As a result it provides also a testbed for analysing the
intricate interplay between integrability and
superintegrability---a testbed having the advantage of being
realistic and intuitive interaction.

Also, from the viewpoint of a RF dynamics enthusiasts we see here
a curious (and encouraging) example when the RF dynamics---exotic
as it may be---behaves `better' than the original problem as it is
always superintegrable.

As a final remark we note that the new results reported here do
not contradict existing classifications of superintegrable models
(see e.g.\ \cite{Evans}) where only first integrals which are low
order polynomials in momenta are described.

\section*{Acknowledgments} This research was partially supported by
PRIN SINTESI. Two of us (AK and VSG) are grateful to Physics
Department of the University of Salerno for the kind hospitality
and acknowledge the partial support by the Italian Istituto
Nazionale di Fisica Nucleare and the National Science Foundation
of Bulgaria, contract No. F-1410.

\end{document}